\documentclass[12pt]{article}
\usepackage{amsfonts,amsmath,amssymb,amscd,mathrsfs}
\usepackage{chicago}
\usepackage{breakcites}
\usepackage[breaklinks=true]{hyperref}

\title{Scope of the action principle}

\author{Ward Struyve\footnote{Institute for Theoretical Physics, KU Leuven, Belgium}$^{*}$\footnote{Centre for Logic and Philosophy of Science, KU Leuven, Belgium}  }

\def\de{\delta}

\def\na{\nabla}
\def\lam{\lambda}

\def\pa{\partial}

\def\ii{\textrm i}
\def\ee{\textrm e}

\def\vp{\varphi}

\newcommand{\be}{\begin{equation}}
\newcommand{\en}{\end{equation}}
\newcommand{\bi}{\begin{itemize}}
\newcommand{\ei}{\end{itemize}}

\begin{document}

\date{}

\maketitle

\begin{abstract}
\noindent
Laws of motion given in terms of differential equations can not always be derived from an action principle, at least not without introducing auxiliary variables. By allowing auxiliary variables, e.g.\ in the form of Lagrange multipliers, an action is immediately obtained. Here, we consider some ways how this can be done, drawing examples from the literature, and apply this to Bohmian mechanics. We also discuss the possible metaphysical status of these auxiliary variables. A particularly interesting approach brings the theory in the form of a gauge theory, with the auxiliary variables as gauge degrees of freedom.

\end{abstract}

\section{Formulating an action}\label{sec1}
Laws of physics are often presented in the form of an action principle where the physically allowed histories correspond to the extrema of an action $S=\int dt L$, given by a Lagrangian $L$ integrated over time \citeN{lanczos70,yourgrau60}. The advantages of such an action principle are manifold. First, it allows for a compact formulation of those laws. Second, Lagrangians that are local in time admit a Hamiltonian formulation \citeN{dirac64,gitman90,hanson76,henneaux92,sundermeyer82}. Third, Noether's (first) theorem establishes a connection between continuous symmetries of the action and conserved currents \cite{brownhr04a}. 

Here, we consider the question to what extent it is possible to formulate such an action principle. More specifically, given laws of physics in the form of a set of differential equations, local in time, can these be derived as the Euler--Lagrange equations of some Lagrangian? Consider for example the discrete case, where the dynamics concerns a configuration $q=(q_1,\dots,q_N) \in {\mathbb{R}}^N$ given by differential equations{\footnote{Higher time derivatives could be considered, but we will not do so for notational simplicity.}}  
\be
f_i(q,\dot q, \ddot q,t )=0, \qquad i=1\dots M. 
\label{1}
\en
Can these be derived from a Lagrangian $L(q,\dot q,t)$?{\footnote{Higher time derivatives need not be considered in the Lagrangian since the equations of motion \eqref{1} only depend on time derivatives of $q$ up to second order.}} This is known as the inverse problem of Lagrangian mechanics, see e.g.\ Santilli \citeyear{santilli78}, and the answer is that it depends on the particular form of the equations \eqref{1}. Some can be derived from a Lagrangian, others not, and there is extensive literature on this. However, a crucial assumption in the inverse problem is that the Lagrangian depends only on $t,q$ and $\dot q$. Admitting auxiliary variables trivializes the problem. Namely, introducing Lagrange multipliers $\lambda_i$, $i=1,\dots, M$, the equations of motion \eqref{1} can immediately be derived from the Lagrangian
\be
L_1 = \sum_i \lambda_i f_i
\label{2}
\en
as constraint equations, by varying with respect to the $\lambda_i$.  But in addition to these equations, variation with respect to $q$ yields 
\be
\sum_i \left[ \frac{d^2}{dt^2}\left(\lambda_i \frac{\pa f_i}{\pa \ddot q_k} \right) - \frac{d}{dt} \left(\lambda_i \frac{\pa f_i}{\pa \dot q_k} \right)  +  \lambda_i \frac{\pa f_i}{\pa q_k}\right] =0 , \qquad k=1,\dots,N.
\label{3}
\en
The equations \eqref{1} do not depend on the $\lambda_i$ and can be solved independently. Given a solution $q(t)$, the equations \eqref{3} then form differential equations for the $\lambda_i$. An early reference proposing this general method is Bateman's \citeyear{bateman31}.
 
Other actions exist which significantly simplify the equations of motion for the auxiliary variables, even to the point where their dynamics is completely arbitrary, i.e., unconstrained by the laws of motion. Consider for example the Lagrangian 
\be
 L_2= \frac{1}{2} \sum_i \lambda_i f^2_i .
\label{3.3}
\en
The corresponding Euler--Lagrange equations are
\be
f^2_i = 0,
\label{3.4}
\en
\be
\sum_i \left[ \frac{d^2}{dt^2}\left(\lambda_i f_i\frac{\pa f_i}{\pa \ddot q_k} \right) - \frac{d}{dt} \left(\lambda_i f_i\frac{\pa f_i}{\pa \dot q_k} \right)  +  \lambda_i f_i\frac{\pa f_i}{\pa q_k}\right] =0.
\label{3.5}
\en
Since the first equation implies $f_i=0$, the second equation is automatically satisfied. So in this case, the desired equations of motion are obtained and the auxiliary variables $\lam_i$ are unconstrained. 

Still different Lagrangians could be considered. For example, in the case of the Lagrangian
\be
L_3=   \frac{1}{2}  \lambda \sum_i f^2_i ,
\en
there is only one auxiliary variable, instead of $M$ as in the case of $L_2$, which is again unconstrained by the equations of motion. The same is true for
\be
L_4=   \frac{1}{2}  \ee^\lambda \sum_i f^2_i ,
\en
with the difference that now the extrema of the action are necessarily minima. Namely for any history $(q(t),\lambda(t))$, the action is non-negative, but for  extrema the action is zero. 

Another example is
\be
L_5 =  \sum_i \left( \lambda_{1i} f_i + \frac{\lambda^2_{1i}}{2} \lambda_{2i}\right),
\en
which involves auxiliary variables $\lambda_{1i}$ and $\lambda_{2i}$, $i=1,\dots, M$. The Euler--Lagrange equations are
\be
\sum_i \left[ \frac{d^2}{dt^2}\left(\lambda_{1i} \frac{\pa f_i}{\pa \ddot q_k} \right) - \frac{d}{dt} \left(\lambda_{1i} \frac{\pa f_i}{\pa \dot q_k} \right)  +  \lambda_{1i} \frac{\pa f_i}{\pa q_k}\right] =0,
\label{5}
\en
\be
 f_i + \lambda_{1i} \lambda_{2i} =0,
\label{6}
\en
\be
  \lambda^2_{1i} =0.
\label{7}
\en
Because of the last equation, these equations reduce to
\be
 f_i  =0, \qquad \lambda_{1i} = 0.
\label{8}
\en
So the variables $\lambda_{1i}$ vanish, while the variables $\lambda_{2i}$ are completely free. Yet another possibility is
\be
L_6 =  \sum_i \left( \lambda_{1i} f_i + \frac{\lambda_{1i}\lambda_{1j}}{2} \lambda_{2ij}\right),
\en 
with $\lambda_{2ij}$ symmetric. These alternatives might be useful in the case one wants the action to be symmetric under certain transformations. For example, in the case of a field theory where the equations of motion have a tensorial character such actions could maintain manifest Lorentz invariance.

So by introducing auxiliary variables, actions can immediately be formulated. While such actions allow for a compact formulation of the laws of motion, it is unclear whether they lead to any technical advantage. Perhaps, using Noether's theorem it is easier to find conservation laws, by identifying continuous symmetries of the action. But this should probably be investigated on a case by case basis. 

What is now the metaphysical status of these auxiliary variables? It might be desired that an action should only include dynamical variables which are regarded as physically real. (Penrose expresses such a sentiment, see below.) For example, the variables $q$ could correspond to the positions of particles. Demanding that the action depends only on those variables (and their time-derivatives) excludes the actions proposed above. However, one could assume the variables $\lambda$ to be physically real as well. Since the dynamics for the variables $q$ is unaffected, the empirical content of the theory---insofar as it is derived from the $q$'s---is the same. But it also implies that part of the world remains forever hidden. This part can evolve in a quite non-trivial way, as in the case of the Lagrangian $L_1$, or in a trivial way (with vanishing or unconstrained variables) as in the other cases $L_2$--$L_6$. Interesting examples of the former case, to be discussed in the next section, are the damped harmonic oscillator and the heat equation, where the auxiliary variables correspond to a doubling of the variables, which evolve according to time-reversed laws of motion. Examples of the latter case are gauge theories. Gauge theories contain unphysical variables---the gauge variables---which happen to evolve completely freely and are regarded as mere mathematical artifacts corresponding to different representations of the same physical reality.{\footnote{While the gauge variables are usually considered as mere mathematical artefacts, there have also been arguments to consider them as physically real, for example in relation to the Aharonov-Bohm effect, see Healey \citeyear{healey07} for a detailed discussion.}} However, actions for gauge theories like Yang--Mills theories or general relativity (where the gauge freedom stems from the freedom of coordinate choice) are most naturally and simply formulated in terms of such variables. This may seem puzzling. Penrose \citeyear[491]{penrose04} writes: ``Moreover, the `Maxwell Lagrangian' does not work as a Lagrangian unless it is expressed in terms of a potential, although the value of the potential $A_a$ is not a directly observable quantity. [\dots] Lagrangians for fields are undoubtedly extremely useful as mathematical devices, and they enable us to write down large numbers of suggestions for physical theories. But I remain uneasy about relying upon them too strongly in our searches for improved fundamental theories.''. Given a dynamics that is derived from a Lagrangian, it has become standard to identify those variables that evolve freely as the gauge variables \cite{dirac64,gitman90,hanson76,henneaux92,sundermeyer82}. As such, the theories described by the Lagrangians $L_2$--$L_6$ count as gauge theories. So in this sense, given a theory expressed in terms of differential equations, it can always be derived from an action principle by turning it into a gauge theory. The action then contains unphysical degrees of freedom (the gauge variables), just as the actions of Yang--Mills theories and general relativity.
 
In the next section, we present examples of actions that use auxiliary variables, often in the form of multipliers. In section \ref{ham}, we present the Hamiltonian formulation for the Lagrangian $L_3$. In sections \ref{bohmian} and \ref{hambohm}, we respectively present an action for Bohmian mechanics, which necessarily involves auxiliary variables, and discuss its Hamiltonian formulation. We conclude in section \ref{conclusion}.

\section{Examples from the literature}
The use of multipliers as in \eqref{2} is a familiar practice in physics. An early example is that of the damped harmonic oscillator, for which the equation of motion reads
\be
\ddot x + 2k \dot x + n^2 x = 0,
\label{8.1}
\en
with $k$ and $n$ constants. Bateman \citeyear{bateman31} considers the Lagrangian
\be
L_7 = y (\ddot x + 2k \dot x + n^2 x),
\label{8.11}
\en
where $y$ is a new variable acting as a Lagrange multiplier. The resulting Euler--Lagrange equations are \eqref{8.1} together with the time-reversed equation for $y$ 
\be
\ddot y - 2k \dot y + n^2 y = 0.
\label{8.2}
\en
So the dynamics for $x$ and $y$ are decoupled and despite the special role of $y$ as multiplier in the action, their dynamics is dual under time-reversal. This duality could also be introduced at the level of the Lagrangian by adding the total time derivative 
\be
-\frac{d}{dt}(y\dot x + k y x)
\en
to the Lagrangian $L_7$ \cite[298]{bateman31,morse58}, resulting in
\be
L_8 = -\dot y \dot x   + k (y \dot x - \dot y x)  + n^2 yx.
\label{8.3}
\en
Such an operation does not affect the action and hence neither  the equations of motion. The variable $y$ no longer appears as a multiplier, but on par with the variable $x$. The role of $x$ and $y$ in the Lagrangian can of course be interchanged, so that $x$ acts as a multiplier:
\be
L_9 = x (\ddot y - 2k \dot y + n^2 y).
\en
This example illustrates that there is a great variety in the role the auxiliary variable can play in the Lagrangian: from Lagrange multiplier to ordinary dynamical variable. (Still other Lagrangians exist which give the equation of motion \eqref{8.1}, also ones that are inequivalent, in the sense that they do not differ merely by a total time derivative. Bateman even gives an example of a Lagrangian which does not employ auxiliary variables, but which is explicitly time dependent.) The Lagrangian for the heat equation is of a form similar to \eqref{8.3}, with an auxiliary field which satisfies the time-reversed dynamics compared to that of the heat field \cite[313]{morse58}. More examples of this type can be found in Ibragimov and Kolsrud \citeyear{ibragimov04}. The Lagrangian for the Schr\"odinger equation is also of a similar form, but with the complex conjugate $\psi^*$ (rather than a new field) in the role of the dual field \cite[314]{morse58}.

Another example is the generally covariant action proposed independently by Rosen \citeyear{rosen66} and Sorkin \citeyear{sorkin02}:
\be
L_{10} = \int d^3 x \sqrt{-g} \lambda^{\mu \nu \rho \sigma} R_{\mu \nu \rho \sigma},
\en
where $g_{\mu \nu}$ is the Lorentzian space-time metric, $R_{\mu \nu \rho \sigma}$ is the Riemann curvature tensor and $\lambda^{\mu \nu \rho \sigma}$ are multipliers (which satisfy the same symmetries as the curvature tensor). The corresponding equations of motion are 
\be
 R_{\mu \nu \rho \sigma} = 0,
\label{riem}
\en
together with equations of motion for $\lambda^{\mu \nu \rho \sigma}$. (The equation \eqref{riem} implies that the metric equals the Minkowski metric, up to space-time diffeomorphisms.) This  Lagrangian $L_{10}$ was used in the debate on the meaning of general covariance in general relativity, see Pitts \citeyear{pitts06} for a detailed discussion.

Our main examples, however, are given by gauge theories. As mentioned in the introduction, gauge theories involve variables that evolve completely freely. Consider for example Maxwell's theory for electromagnetism in the absence of charges. The Lagrangian is usually taken to be{\footnote{Throughout the paper units are used so that $c=\hbar=1$.}}
\be
L_{11} = -  \frac{1}{4}\int d^3 x F^{\mu \nu} F_{\mu \nu} ,
\label{m0}
\en
where $F_{\mu \nu} = \pa_\mu A_\nu - \pa_\nu A_\mu$ is the electromagnetic field tensor and $A^\mu =(\vp, {\bf A})$ is the vector potential, leading to the Maxwell equations $\pa _\mu F^{\mu \nu} =0$. The theory has a gauge symmetry, given by 
\be
A^\mu \to A^\mu + \pa^\mu \theta, 
\label{m0.01}
\en
with $\theta$ an arbitrary function of space and time. This symmetry maps solutions of the Maxwell equations to solutions. It is often regarded as an unphysical symmetry which merely connects different mathematical representations of the same physical history (see Healey \citeyear{healey07} for dissenting views). Assuming that the field vanishes sufficiently fast at spatial infinity, the Helmholtz decomposition ${\bf A} = {\bf A}^T + {\bf A}^L$ can be applied \cite{griffithsdj99}, where 
\be
{\bf A}^T = {\bf A} - {\boldsymbol \nabla} \frac{1}{\nabla^2} {\boldsymbol \nabla}  \cdot {\bf A} ,\qquad {\bf A}^L = {\boldsymbol \nabla} \frac{1}{\nabla^2} {\boldsymbol \nabla} \cdot {\bf A}
\en
are the transverse and longitudinal part of the vector potential (${\boldsymbol \nabla} \cdot {\bf A}^T = {\bf 0}$ and ${\boldsymbol \nabla} \times {\bf A}^L = {\bf 0}$) and $\nabla^{-2} f({\bf x})= - \int d^3 y f({\bf y})/4\pi|{\bf x}-{\bf y}|$. The action can be written as
\be
L_{12}= \frac{1}{2}\int d^3 x\left[ \dot {\bf A}^T \cdot \dot {\bf A}^T +{\bf A}^T \cdot \nabla^2  {\bf A}^T + \dot {\bf A}^L \cdot \dot {\bf A}^L - 2 \vp {\boldsymbol \nabla} \cdot \dot {\bf A}^L - \vp \nabla^2 \vp \right] . 
\en
In terms of these variables, the Maxwell equations are{\footnote{For varying the action, one can express the potential in terms of Fourier modes.}}
\be
 \ddot {\bf A}^T  -  \nabla^2  {\bf A}^T = {\bf 0},
\en
\be
 \ddot {\bf A}^L  + {\boldsymbol \nabla} \dot \vp = {\bf 0},
\label{m2}
\en
\be
{\boldsymbol \nabla} \cdot  \dot {\bf A}^L + \nabla^2 \vp  = 0.
\label{m3}
\en
The second equation \eqref{m2} follows from \eqref{m3} (by applying $\frac{\partial}{\partial t} {\boldsymbol \nabla} \nabla^{-2} $) 
and is hence redundant. The dynamics of ${\bf A}^T$ is decoupled from that of $\vp$ and ${\bf A}^L$. The dynamics of the latter is such that any function $\vp$ or any ${\bf A}^L$ is allowed with the only constraint that they are mutually correlated by \eqref{m3}. Put differently, we can have any field ${\bf A}^L$ as a solution, provided $\vp$ is determined by \eqref{m3} (which itself does not pose restrictions on ${\bf A}^L$). Conversely, any function $\vp$ is allowed with ${\bf A}^L$ determined by \eqref{m3}. This freedom in time evolution is also clear from the gauge symmetry \eqref{m0.01}, since a gauge transformation does not affect ${\bf A}^T$ but only $\vp$ and ${\bf A}^L$. Because the evolution of $\vp$ and ${\bf A}^L$ is arbitrary and does not affect the evolution of ${\bf A}^T$, they are traditionally considered unphysical variables. This arbitrary time evolution was also encountered for auxiliary variables in the Lagrangians $L_2$--$L_6$ of the previous section, making the corresponding theories gauge theories in the light of the present discussion. 

In the present case, the gauge degrees of freedom can easily be dismissed, by considering just the transverse part of the potential and keeping only the first two terms in the Lagrangian $L_{12}$. However, this comes at the price of losing manifest Lorentz invariance. Moreover, for more complicated theories like non-Abelian Yang--Mills theories or general relativity, it becomes very difficult (if at all possible globally) to express the action or equations of motion in terms of gauge invariant quantities.

There are also actions in terms of the electric and magnetic field (or the field strength), rather than the potentials, but these also involve auxiliary fields acting as Lagrange multipliers \cite{vollick17}. Consider for example the Lagrangian:
\be
L_{13} = -  \int d^3 x \left( \frac{1}{4} F^{\mu \nu} F_{\mu \nu}  + A_\nu \pa_\mu F^{\mu \nu} \right) ,
\en
where $A_\mu$ and $F_{\mu \nu}$ are treated as independent fields \cite{infeld54,schwinger51}. Clearly, the field $A_\mu$ acts as a Lagrange multiplier, implying
\be
\pa_\mu F^{\mu \nu} =0.
\en
Variation with respect to $F_{\mu \nu}$ leads to the familiar relation{\footnote{The equation of motion \eqref{m10} yields $F_{\mu \nu}$ in terms of $A_\mu$. This is an example of what is sometimes called an {\em auxiliary variable} \cite{pons10}. (This technical notion of ``auxiliary variable'' should be contrasted with the colloquial notion we have been using in the rest of the paper.) That is a variable whose variation in the action leads to an equation of motion that allows to solve that variable in terms of the other fields. Such variables can be introduced for simplification. They can also be eliminated without changing the dynamics of the other fields. On the level of the action, this can be done by simply substituting the expression for that variable (in terms of the other fields) into the action. In the present case, such an elimination results in the Lagrangian $L_{11}$. A further reduction is possible since also $\phi$ is an auxiliary variable (in the technical sense), cf.\ \eqref{m3}. Its elimination yields the Lagrangian for just the transverse potential.}}
\be
F_{\mu \nu} = \pa_\mu A_\nu - \pa_\nu A_\mu.
\label{m10}
\en
So the latter is not assumed, as in eq.~\eqref{m0}, but is instead derived as one of the equations of motion. A similar situation arises in general relativity. The Einstein equations can be derived from the Einstein--Hilbert action by varying with respect to the metric $g_{\mu \nu}$ (whose components also contain gauge degrees of freedom in the sense explained above). An alternative way is via the Palatini method which treats the connection and metric as independent fields \cite{ferraris82,misner17,wald84}.

\section{Hamiltonian formulation}\label{ham}
The forms of the Lagrangians $L_1$--$L_6$ allow for the possibility of a Hamiltonian formulation with the usual methods. Consider for example $L_3$ and take $f_i=f_i(q,\dot q,t)$, so that the Lagrangian contains time derivatives of $q$ only up to order one. Then the conjugate momenta are
\be
p_k = \frac{\pa L}{\pa \dot q_k} = \lam \sum_i f_i \frac{\pa f_i}{\pa \dot q_k} ,
\label{h1.1}
\en
\be
\pi = \frac{\pa L}{\pa \dot \lam} = 0.
\label{h1.2}
\en 
Because of the latter equation, these relations are not invertible to yield the velocities in terms of the phase-space variables. This means that we have to resort to the theory of constrained dynamics \cite{dirac64,gitman90,hanson76,henneaux92,sundermeyer82}. An immediate primary constraint is 
\be
\pi =0.
\label{h2}
\en
For simplicity, we assume that there are no further primary constraints, so that the $\dot q_k$ can be expressed in terms of the phase-space variables, i.e., there are functions $v_k(q,p,\lam,t)$ such that the relations \eqref{h1.1} can be inverted to yield
\be
\dot q_k = v_k(q,p,\lam,t).
\en 
The canonical Hamiltonian is
\begin{align}
H_c  &= \sum_k p_k \dot q_k + \pi \dot \lam - L_3 \nonumber\\
 &= \sum_k p_k v_k - \frac{1}{2}\lam \sum_i f^2_i(q,v,t).
\end{align}
The total Hamiltonian is 
\be
H_T = H_c + u \pi,
\en
where $u$ is an arbitrary function of the phase-space variables. The corresponding Hamilton's equations, together with the constraint \eqref{h2}, give the equations of motion in phase-space. With the constraint taken into account, Hamilton's equations are
\begin{align}
\dot q_k &= v_k,\\
\dot p_k &= \lam \sum_{i} f_i \frac{\pa f_i}{\pa  q_k} ,\\
\dot \lam &= u,\\
\dot \pi &= \frac{1}{2} \sum_i f^2_i
\end{align}
(where the definitions of the momenta were used to simplify the expressions). Since $u$ was an arbitrary function, we have again (as of course we should) that the evolution of $\lam$ is arbitrary. By the last equation, the constraint $\pi =0$ further implies that $f_i(q,v(q,p,\lam,t),t)=0$. These are the secondary constraints. 
Considering the definitions \eqref{h1.1}  of the momenta $p_k$, it must be the case that for a solution of the equations of motion, they are zero, i.e., $p_k=0$. So these constraints $f_i=0$ must amount to $p_k =0$. Using these relations, the equations of motion can be simplified to
\begin{align}
\dot q_k &= v_k (q,0,\lam,t),\label{h20}\\
\dot p_k &= 0,\\
\dot \lam &= u,\\
\dot \pi &= 0.
\end{align}
In the first relation \eqref{h20}, there is a dependence on $\lam$. However, because of \eqref{h1.1}, the $v_k$ depend on $\lam$ only through $p_k/ \lam$ and hence with $p_k =0$, this implies that there is no $\lam$-dependency of $v_k$. So equation \eqref{h20} amounts to the equations $f_i(q,\dot q,t)$ expressed in the form $\dot q_k = g_k(q,t)$ for certain functions $g_k$. 

Actually, given equations of motion of the form $\dot q_k = g_k(q,t)$, the Hamiltonian formulation can be done with the Hamiltonian
\be
H = \sum_k p_k  g_k(q,t),
\en
together with the constraints $p_k=0$. The equations of motion then immediately reduce to $\dot q_k = g_k(q,t)$ and $p_k=0$. 

The Hamiltonian is a constant of the motion, but in this case it is a trivial one as it vanishes along a solution.

In the next section, we will provide an example of this Hamiltonian formulation for Bohmian mechanics.

\section{Application to Bohmian mechanics}\label{bohmian}
Bohmian mechanics concerns the motion of point-particles whose velocity depends on the wave function \cite{bohm93,duerr12,duerr09,holland93b}. The wave function satisfies the usual Schr\"odinger equation, whereas the particles satisfy the so-called guidance equations. In the case of a single particle (which we consider here for mere notational simplicity), denoting the position of the particle at time $t$ by ${\bf X}(t)$ and the wave function by $\psi({\bf x},t)$, the dynamics is
\be
\dot {\bf X}(t) = {\bf v}^\psi({\bf X}(t),t), 
\label{10}
\en
where
\be
{\bf v}^\psi({\bf x},t) = \frac{1}{m} {\textrm{Im}} \left(\frac{ {\boldsymbol \nabla} \psi({\bf x},t)}{\psi({\bf x},t)} \right) ,
\en
\be
\ii \frac{\pa \psi({\bf x},t)}{\pa t} = -\frac{1}{2m} \nabla^2\psi({\bf x},t) + V({\bf x}) \psi({\bf x},t).
\en

The Schr\"odinger equation can be derived from the Lagrangian $L_S=\int d^3x {\mathcal L}_S$, with ${\mathcal L}_S$ the Lagrangian density given by 
\be
{\mathcal L}_S = \frac{1}{2} \psi^* \left(\ii \frac{\pa \psi}{\pa t} + \frac{1}{2m} \nabla^2\psi - V \psi \right) + {\textrm{c.c.}}
\en 
While the Euler--Lagrange equations can be found by formally treating $\psi$ and $\psi^*$ as independent fields, this Lagrangian is better viewed as a function of the real and imaginary part of $\psi$, which {\em are} independent \cite{brownhr04b}.

The Bohmian dynamics cannot be derived from a Lagrangian that depends only on ${\bf X}$ and $\psi$. So a Lagrangian requires the introduction of auxiliary variables. There have been attempts by Squires \citeyear{squires94b} and Holland \citeyear{holland01a,holland06,holland20} to write down such a Lagrangian. However, these proposals do not exactly recover the Bohmian dynamics, but rather some generalized dynamics, for which the Bohmian trajectories are only a subset of the possible allowed trajectories.

Squires \citeyear{squires94b} considers the Lagrangian density{\footnote{Squires considers a further addition to this Lagrangian given by a constant $k$ times the standard non-relativistic particle Lagrangian. Only the special case $k=0$ is presented here.}}
\be
{\mathcal L}_{\textrm{Sq}} = {\mathcal L}_S + {\boldsymbol \lambda} \cdot \left(\dot {\bf X} - {\bf v}^\psi \right) \delta({\bf x} - {\bf X}).
\en
Variation of the action with respect to ${\boldsymbol \lambda}$ yields the guidance equation. Variation with respect to $\psi$ and ${\bf X}$ respectively gives
\be
\ii \frac{\pa \psi}{\pa t} = -\frac{1}{2m} \nabla^2\psi + V \psi - \frac{\ii}{2m \psi^* } {\boldsymbol \lambda} \cdot {\boldsymbol \na} \delta({\bf x} - {\bf X}),
\en
\be
\frac{d {\boldsymbol \lambda}}{dt} + {\boldsymbol \na} \left[{\boldsymbol \lambda} \cdot {\bf v}^\psi({\bf x}) \right] \big|_{{\bf x} = {\bf X}} = {\bf 0}.
\en
So while the guidance equation is obtained, the Schr\"odinger equation gets an extraneous ${\boldsymbol \lambda}$-dependent contribution. Only in the case ${\boldsymbol \lambda}={\bf 0}$ the Bohmian dynamics is recovered. For ${\boldsymbol \lambda} \neq {\bf 0}$ the Schr\"odinger equation is not satisfied.

Holland \citeyear{holland20} considers{\footnote{Holland considers a similar Lagrangian in \cite{holland01a,holland06} using a different parameterization of the wave function.}}
\be
{\mathcal L}_{\textrm{Ho}} = \frac{1}{2} u^* \left(\ii \frac{\pa \psi}{\pa t} + \frac{1}{2m} \nabla^2\psi - V \psi \right) + {\textrm{c.c.}} + \left( \frac{1}{2} m \dot {\bf X} \cdot \dot {\bf X} - V - Q^\psi \right)  \delta({\bf x} - {\bf X}),
\en 
where 
\be
Q^\psi({\bf x},t) = - \frac{1}{2m} \frac{\nabla^2|\psi({\bf x},t)|}{|\psi({\bf x},t)|}
\en
is the quantum potential. In this case the complex field $u$ is introduced as the Lagrange multiplier. As a result, variation with respect to $u$ gives the Schr\"odinger equation, while variation with respect to respectively ${\bf X}$ and $\psi$ yields
\be
m \ddot {\bf X} = - {\boldsymbol \nabla} (V({\bf x})+Q^\psi({\bf x},t))\big|_{{\bf x} = {\bf X}} ,
\label{20}
\en
\be
\ii \frac{\pa u}{\pa t} = -\frac{1}{2m} \nabla^2u + V u +2 \frac{\pa Q^\psi}{\pa \psi^*}\bigg|_{{\bf x} = {\bf X}}.
\en 
While this action yields the desired Schr\"odinger equation, it does not yield the guidance equation. The Newtonian-like equation \eqref{20} follows from the Bohmian dynamics by taking the time derivative of the guidance equation (using also the Schr\"odinger equation). But \eqref{20} also allows for non-Bohmian solutions, where the guidance equation does not hold, i.e., where the velocity is different from ${\bf v}^\psi$. Holland considers the guidance equation as an extra constraint on the dynamics (which can be imposed at an initial time). This is necessary for empirical adequacy \cite{colin13,goldstein15}.

As explained in the introduction, Lagrange multipliers can be introduced to enforce both the guidance equation and the Schr\"odinger equation. But as can readily be checked, the following Lagrangian density already yields the Bohmian dynamics
\be
{\mathcal L}_{\textrm{B}} = {\mathcal L}_S +  \frac{\lambda}{2}  \left(\dot {\bf X} - {\bf v}^\psi \right)\cdot \left(\dot {\bf X} - {\bf v}^\psi \right) \delta({\bf x} - {\bf X}) .
\label{22}
\en
So there is no need to introduce Lagrange multipliers to enforce the Schr\"odinger equation. The variable $\lambda$ evolves freely and is considered a gauge variable in the context of the theory of constrained dynamics. 

Before turning to the Hamiltonian formulation, let us have a look at possible Noether currents. The action $S_{\textrm{B}} = \int dt d^3x {\mathcal L}_{\textrm{B}}$ is invariant under Galilean transformations (with the usual transformations for ${\bf X}$ and $\psi$ \cite{duerr09}. However, the Noether currents corresponding to space and time translations, rotations and boosts, are just the usual currents associated to the Schr\"odinger equation, because the $\lambda$-dependent part of these currents is proportional to $\dot {\bf X} - {\bf v}^\psi$ and hence vanishes for a solution.

\section{Hamiltonian formulation of Bohmian mechanics}\label{hambohm}
The Hamiltonian formulation outlined in section \ref{ham} can now be applied to the Bohmian Lagrangian \eqref{22}, with only an extra complication arising from the treatment of the Schr\"odinger part, which can however be found elsewhere, e.g.\ \cite{gergely02}.

Writing $\psi = \psi_1 + \ii \psi_2$, with $\psi_1$ and $\psi_2$ real, the conjugate momenta are
\be
\pi_1 = \frac{\de L_{\textrm{B}} }{\de \dot \psi_1 } = \psi_2 , \quad \pi_2 = \frac{\de L_{\textrm{B}} }{\de \dot \psi_2} =- \psi_1 ,
\label{23}
\en
\be
P_i = \frac{\pa L_{\textrm{B}} }{\pa \dot X_i} = \lambda \left[\dot X_i - v^\psi_i({\bf X}) \right] , \quad  \pi =  \frac{\pa L_{\textrm{B}} }{\pa \dot \lambda} = 0 .  
\label{24}
\en
As in section \ref{ham}, we are dealing with a constrained dynamics because these relations are not invertible to yield the velocities in terms of the phase-space variables.  The non-invertible relations yield the primary constraints
\be
\pi_1 - \psi_2 =0, \qquad \pi_2 + \psi_1 =0, \qquad \pi = 0.
\label{25}
\en
The canonical Hamiltonian is
\begin{align}
H_c &= \int d^3 x \left(\pi_1 \dot \psi_1 + \pi_2 \dot \psi_2 \right) + {\bf P} \cdot \dot {\bf X} + \pi \dot \lam  - L_{\textrm{B}}  \nonumber\\
&=\int d^3 x \psi^*{\widehat H} \psi  + {\bf P} \cdot {\bf v}^\psi({\bf X}) + \frac{1}{2\lam} {\bf P} \cdot {\bf P}  ,
\end{align}
where 
\be
{\widehat H} =  -\frac{1}{2m} \nabla^2  + V .
\en

The total Hamiltonian is
\be
H_T = H_c + \int d^3 x \left[ u_1 (\pi_1 - \psi_2) +u_2 (\pi_2 + \psi_1) \right] + w  \pi,
\en
where $u_1({\bf x})$, $u_2({\bf x})$ and $w$ are arbitrary functions of the phase-space variables. This Hamiltonian determines the equations of motion through the usual Hamilton's equations, together with the constraints \eqref{25}. It can be simplified by deriving the secondary constraints which follow from the fact that the primary constraints need to be preserved in time. Preservation of $\pi = 0$ leads to the constraint ${\bf P} = {\bf 0}$. Using the latter, preservation of the other constraints leads to the constraints
\be
u_1 ={\widehat H}\psi_2,  \qquad u_2 = -{\widehat H}\psi_1.
\en
So the arbitrary functions $u_1$ and $u_2$ get determined by the equations of motion and can be substituted in $H_T$ to yield{\footnote{The first part of the Hamiltonian is that of the wave function and could also be written as $\int d^3 x \psi^* {\widehat H} \psi$ provided the Dirac bracket is used for the Hamilton's equations rather than the Poisson bracket.}} 
\be
H_1 = \int d^3 x \left[ \pi_1 {\widehat H} \psi_2  -  \pi_2 {\widehat H} \psi_1     \right] +  {\bf P} \cdot {\bf v}^\psi({\bf X}) + \frac{1}{2\lam} {\bf P} \cdot {\bf P} + w \pi.
\en
The corresponding Hamilton's equations are now (using the constraint ${\bf P} = {\bf 0}$),
\be
\dot \psi = -\ii {\widehat H} \psi, \qquad \dot \pi_1  =  {\widehat H} \pi_2, \qquad \dot \pi_2  =  - {\widehat H} \pi_1, 
\label{98}
\en
\be
 \dot {\bf X} = {\bf v}^\psi({\bf X}), \qquad \dot {\bf P} = {\bf 0} , 
\label{99}
\en
\be
\dot \lam = w ,\qquad \dot \pi = 0.
\label{100}
\en
So these are the equations of motion of Bohmian mechanics together with equations for $\lam$ and the conjugate momenta. Since $w$ is an arbitrary function, the evolution of $\lam$ is again arbitrary. 

The variables $\lam$ and $\pi$ do not enter the equations of motion for the other canonical variables. Their presence merely implies the constraint ${\bf P} = {\bf 0}$. Keeping the latter constraint, the variables $\lam$ and $\pi$ can be removed by considering the Hamiltonian
\be
H_2  =  \int d^3 x \left[ \pi_1 {\widehat H} \psi_2  -  \pi_2 {\widehat H} \psi_1     \right]   + {\bf P} \cdot {\bf v}^\psi({\bf X}).
\en
The corresponding Hamilton's equation are again \eqref{98} and \eqref{99}, taking into account ${\bf P} = {\bf 0}$. This is the Hamiltonian formulation of Bohmian mechanics proposed by Vollick \citeyear{vollick19}.

Note that, in line with what we said earlier about the Noether currents for this case, the conserved energy (i.e., Hamiltonian) is just that of the wave function, since the particle-dependent part in $H_2$ vanishes along a solution.

\section{Conclusion}\label{conclusion}
Laws of motion given by differential equations can always be derived from an action,  at least if auxiliary variables are allowed. Moreover, these extra variables can be introduced in such a way that they would be regarded as gauge variables according to the usual approach to gauge theories. So these actions are similar in that respect to theories like Yang--Mills theories or general relativity. However, while it is easy to introduce auxiliary variables as gauge variables, it tends to be very hard to eliminate the gauge for theories like Yang--Mills theories and general relativity on the level of the action as well as the dynamics.

\section{Acknowledgments}
This work is supported by the Research Foundation Flanders (Fonds Wetenschappelijk Onderzoek, FWO), Grant No.\ G0C3322N. It is a pleasure to thank Thibaut Demaerel, Christian Maes, Sylvia Wenmackers, and two anonymous referees, for useful comments and discussions.

\end{document}